# Hierarchy of Graphene Wrinkles Induced by Thermal Strain Engineering

Lan Meng[1], Ying Su[1], Dechao Geng[2], Gui Yu[2], Yunqi Liu[2], Rui-Fen Dou[1], Jia-Cai Nie[1], and Lin He[1,*]

[1] Department of Physics, Beijing Normal University, Beijing, 100875, People's Republic of China

[2] Beijing National Laboratory for Molecular Sciences, Key Laboratory of Organic Solids, Institute of Chemistry, Chinese Academy of Sciences, Beijing, 100190, People's Republic of China

**Here we study hierarchy of graphene wrinkles induced by thermal strain engineering and demonstrate that the wrinkling hierarchy can be accounted for by the wrinklon theory. We derive an equation $\lambda = (ky)^{0.5}$ explaining evolution of wrinkling wavelength $\lambda$ with the distance to the edge $y$ observed in our experiment by considering both bending energy and stretching energy of the graphene flakes. The prefactor $k$ in the equation is determined to be about 55 nm. Our experimental result indicates that the classical membrane behavior of graphene persists down to about 100 nm of the wrinkling wavelength.**

Graphene, being a one atom thick membrane, is always wrinkled to certain degree [1,2]. Lattice distortions of the graphene wrinkles couple to the electrons in the same way as effective electric and magnetic fields [3-15], which opens the way for novel applications that are unique for graphene. Importantly, wrinkles in graphene are more than just electronic curiosity, since the atomic thickness makes graphene the ultimate thin film for exploring membrane physics and mechanics [16-22]. These studies show that the classical membrane nature of graphene persists for wavelengths of wrinkles of several hundreds nanometers. In a recent work, a theory based on wrinklon [23], a localized transition region in which two wrinkles with different wavelengths merge, was proposed to account for the wrinkle formation and a universal self-similar hierarchy of wrinkles in different thin films [24-27]. According to this theory, the dependency of the average wavelength $\lambda$ of the hierarchical wrinkles on the distance to the constrained edge $y$ can be described by a simple power law, $\lambda \sim y^m$, and this power law is validated for films with thickness spanning about seven orders of magnitude (here $m$ = 2/3 or 1/2 depending on material properties) [23]. The hierarchical patterns of a suspended graphene bilayer are also demonstrated to follow this power law with $m$ = 1/2, which suggests that the wrinklon theory may be also valid for wrinkles in graphene systems [23].

In spite of this appealing conception, a lack of equivalent advances in systematically measuring the wrinkling hierarchy in graphene systems, to some extent, makes the wrinklon theory suspensive in this ultimate thin film. For example, experimental realization of a single graphene wrinklon is still a very big challenge that difficult to overcome because of the one-atom thickness of graphene. In this Letter, we present a systematic study of the hierarchical patterns in graphene monolayer grown on liquid copper surfaces [28,29] and direct observe the single graphene wrinklon in the flakes. The formation mechanism for quasi-one-dimensional graphene wrinkles on liquid copper surfaces is carefully explored. An equation $\lambda = (ky)^{0.5}$ [23] was derived to describe the hierarchy of graphene wrinkles.

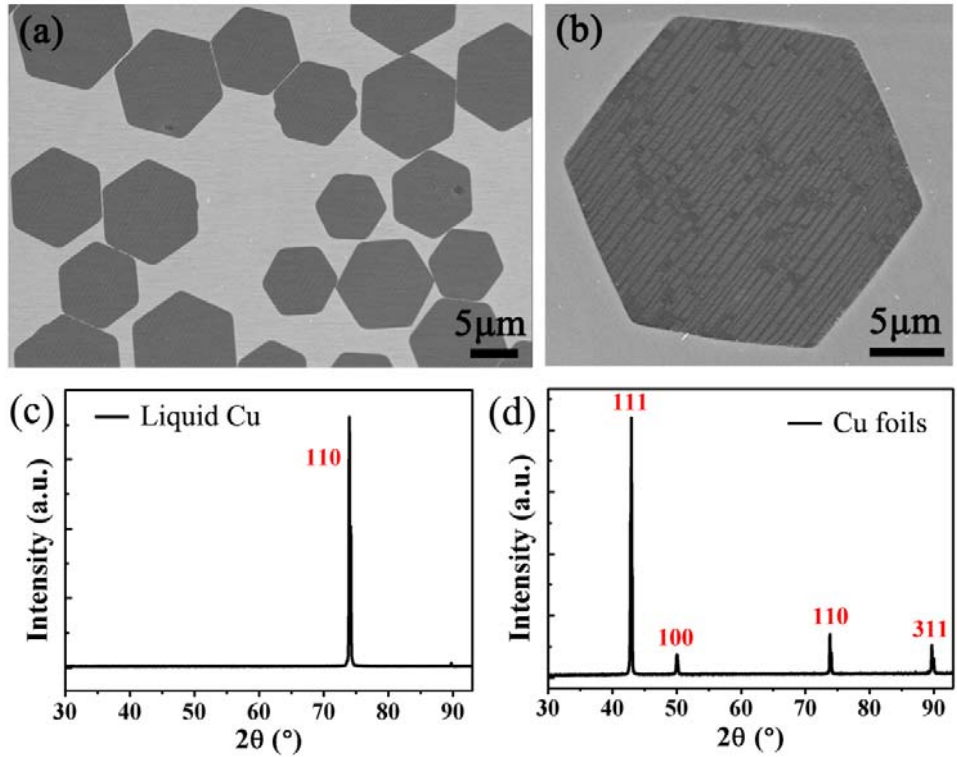

FIG. 1. (a) A typical scanning electron microscopy (SEM) image showing well-dispersed single-crystalline hexagonal graphene flakes on the surface of a Cu substrate. All the graphene flakes show quasi-one-dimensional wrinkles and the wrinkles in different graphene flakes are parallel. (b) A typical SEM image of a hexagonal graphene flake. (c) XRD spectrum of graphene grown on liquid Cu only shows one lattice facet, Cu(110). (d) XRD spectrum of graphene grown on Cu foils shows four different Cu lattice facets.

The graphene monolayer was grown on liquid copper surface by chemical vapor deposition (CVD), as reported in previous papers [28,29]. The approach involves the formation of liquid Cu phase on W substrate at the growth temperature above melting point of Cu, ~1080 °C. Using liquid Cu eliminates the grain boundaries found in solid Cu and produces single-layered, single-crystalline, hexagonal graphene flakes, as shown in Figure 1(a) and 1(b). During the cooling process, mismatch of thermal expansion coefficients between graphene and the substrate results in the formation of wrinkles [6,13,14,16,17]. In our experiment, we usually observe quasi-one-dimensional wrinkles in the hexagonal graphene flakes, and the wrinkles in different graphene flakes on a Cu surface are usually parallel. The parallel wrinkles in graphene flakes grown on liquid copper surface indicate that there is an effective uniaxial force, which affects the wrinkle formation. This wrinkling pattern is distinct from that of graphene flakes grown on solid Cu foils [30] and other solid metal substrates [31,32]. For graphene grown on solid metal surfaces, the directions of the wrinkles in different graphene flakes are of randomly distributed. This difference suggests that the graphene flakes grown on liquid copper are subjected to an anisotropic compression whereas this uniform anisotropic compression is absent for graphene grown on other polycrystalline metal substrates.

The origin of the anisotropic compression is mainly attributed to the anisotropic surface

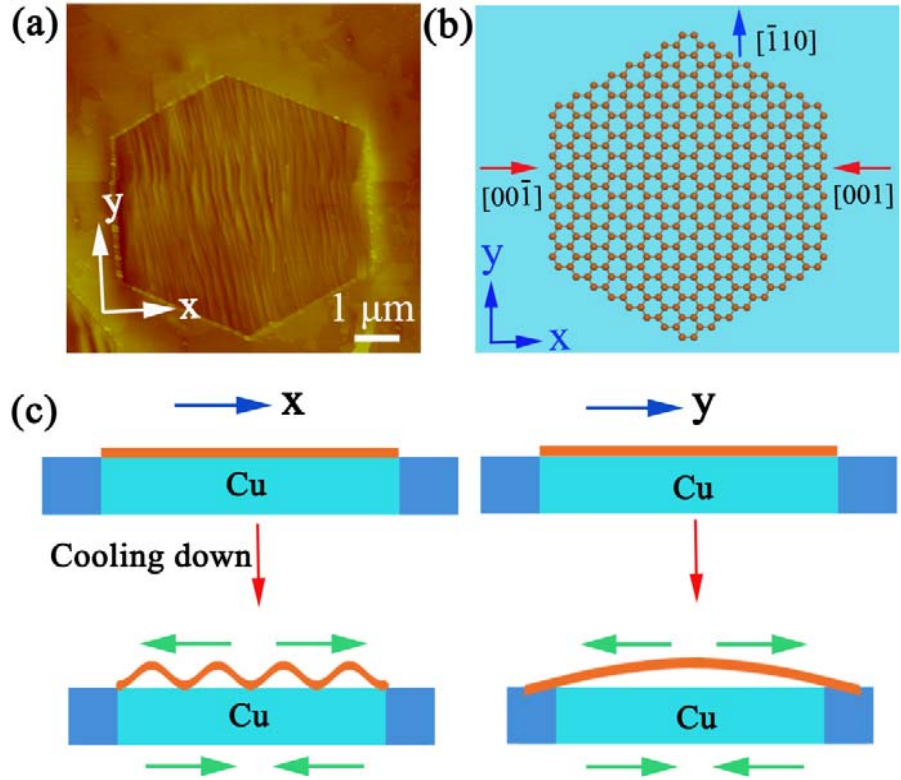

FIG. 2 (color online). (a) An AFM image of a typical hexagonal graphene flake on Cu surface. (b) Schematic diagram showing the top view of a hexagonal graphene flake on Cu(110). (c) Schematic diagrams showing the effect of thermal expansion mismatch on the formation of graphene wrinkles. During the cooling process, the Cu substrate contracts and the graphene expands. Owing to the anisotropy of the surface stress of Cu(110), the graphene wrinkled along the Cu[001] lattice direction (here we defined as *x* axis).

stress of the Cu substrate. Fig. 1(c) shows a typical X-ray diffraction (XRD) spectrum of the Cu underneath the graphene after growth. There is only one crystal facet peak, Cu(110). It indicates that the Cu substrate is a single crystalline that exposes the {110} plane [33]. The surface stress in Cu[001] direction is larger than that in [-110] direction on the Cu(110) surface [34,35]. Such an anisotropy may lead to anisotropic shrinkage of the Cu(110) surface during the cooling process, which plays a vital role in the formation of the uni-directional parallel graphene wrinkles. On the contrary, the solid Cu foils expose several different crystal facets rather than the (110) surface, as shown in Figure 1(d). The polycrystalline structure of solid metal foils results in the ramdomly distribution of the graphene wrinkles along different directions [30-32].

Figure 2(a) shows a typical atomic force microscopy (AFM) image of a hexagonal graphene flake with uni-directional parallel wrinkles. The shrinkage of Cu(110) surface in [001] direction ($x$ direction) is expected to be larger than that in [-110] direction ($y$ direction) [33-35], which gives rise to a larger compression in graphene along [001] direction, as sketched in Fig. 2(b). The stronger compression along the [001] direction induce uni-directional parallel wrinkles along its perpendicular direction, as sketched in Fig. 2(c). Similar to other thin films in a uniaxial strain, the out-of-plane displacement of the hexagonal graphene flakes can be approximated by $z = A\sin(2\pi x/\lambda)\sin(2\pi y/\lambda')$, where $A$ is the amplitude, $\lambda$ and $\lambda'$ are the wavelengths in $x$ and $y$ axes respectively (here $\lambda' \sim W \gg \lambda$ and $W$ is the width of graphene flake) [24,26]. We cannot observe periodic wrinkles along the $y$ direction because the request of least bending energy in the graphene sheets [24].

The above result can be further justified. Because of the formation of uni-directional parallel graphene wrinkles along the direction of $x$ axis, the relative change in length of graphene along $y$ axis can be estimated by $\Delta_y = \int_{T_0}^{T_1} \alpha_G(T) dT$, where $\alpha_G$ is the thermal expansion coefficient of graphene, $T_0$ and $T_1$ are initial and final temperatures during the cooling process, respectively. The relative change in length of graphene with respect to Cu surface along $x$ axis is $\Delta_x = \int_{T_0}^{T_1} [\alpha_G(T) - \alpha_{Cu}(T)] dT$. Here $\alpha_{Cu}$ is the thermal expansion coefficient of Cu surface. During the cooling process, *i.e.*, from about 1100 K to about 300 K, $\alpha_G$ is always much less than $\alpha_{Cu}$ [36-38], and $|\alpha_G - \alpha_{Cu}| > |\alpha_G|$. Then $\Delta_x$ is much larger than $\Delta_y$, therefore, we can observe clear periodic wrinkles only in the $x$ direction but not in the $y$ direction.

According to the wrinklon theory, a single wrinklon was defined as the localized transition zone needed for merging two wrinkles with different wavelengths [23]. To study the behavior of a single wrinklon, Vandeparre, *et al.* constrained two opposite edges of thin plastic sheets by sinusoidal clamps with a wavelength $\lambda$ (amplitude $A$) and $2\lambda$ (amplitude

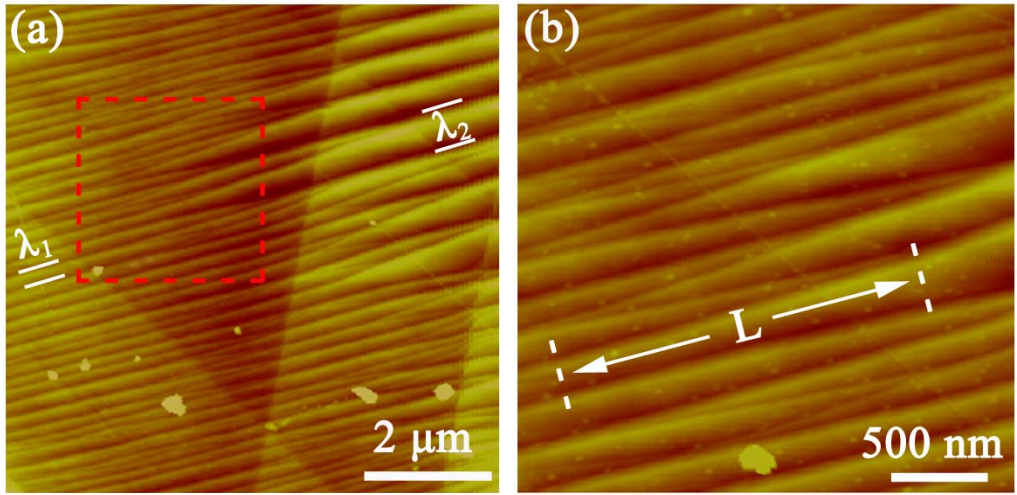

FIG. 3 (color online). (a) AFM image in the middle of a hexagonal graphene flake on Cu surface. It shows morphology of the transition of graphene wrinkles from wavelength $\lambda_1$ to $\lambda_2$. (b) A zoom-in topography in the red frame of (a) shows wrinklons in the localized transition zone where two wrinkles of different wavelengths merge. Here $L$ is the length of a typical wrinklon.

2A), respectively [23]. It is very difficult to carry out the above model experiment in graphene system because of its atomic thickness and the nano-scale wavelength of the wrinkles. However, we can direct observe a single wrinklon in the middle of the hexagonal graphene flakes grown on Cu surface. Figure 3 shows a typical AFM image in the middle of a graphene flake and we observe a transition region where the wavelength of graphene wrinkles increases from $\lambda_1$ to $\lambda_2$. Such a transition induces a distortion of the flake which relaxes over a distance $L$, as shown in Fig. 3(b). Then the variation of the wavelength, $d\lambda/dy$, is of order $\lambda/L$ (here $\lambda = \lambda_2 - \lambda_1$), and the dependency of the average wavelength $\lambda$ of the graphene hierarchical wrinkles on the distance to the edge $y$ can be written as [23]

$$\frac{d\lambda}{dy} \simeq \frac{\lambda}{L}. \qquad (1)$$

The complete hierarchical patterns in graphene systems is expected to be described well by this equation. Figure 4 shows a typical AFM image around a boundary of a hexagonal graphene flake on Cu surface. We can observe obvious wrinkling hierarchy near the edge of graphene. The average wavelength increases with the distance to the edge $y$. All the other graphene flakes on Cu surface also show the universal self-similar hierarchical structures near the constrained boundaries.

Figure 5 summarizes the evolution of the average wavelength $\lambda$ with the distance $y$ to the edges of different graphene flakes. The experimental data in constrained graphene bilayer, as reported in Ref. [23], are also plotted. All the data follow a simple equation $\lambda \approx (ky)^m$ with $k \approx 55$ nm and $m \approx 0.5$. Therefore, our experimental result confirms that the wrinklon theory is also valid for wrinkles in graphene systems.

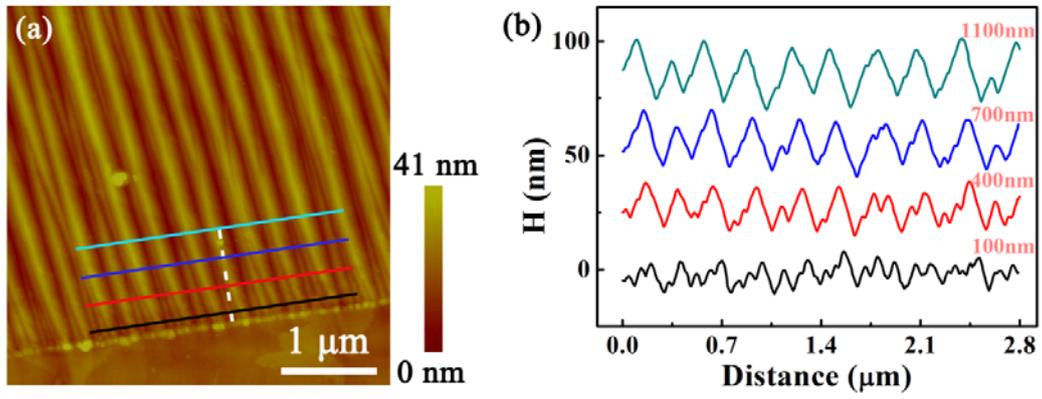

FIG. 4 (color online). (a) A typical AFM image showing the hierarchy wrinkles close to one edge of a hexagonal graphene flake. (b) A series of profile lines parallel to the edge of the graphene flake at different distances, as marked by the lines in (a). The profile lines were vertically offset for clarity.

To understand the parameter $k \approx 55$ nm and $m \sim 0.5$ obtained in our experiment, we further consider the mechanism of the wrinkle formation in graphene system. For graphene, both the bending energy $U_B$ and the stretching energy $U_S$ should be taken into account for the wrinkles formation [39]. Theoretically, the bending energy of the graphene flakes can be expressed as [24]

$$U_B = \frac{1}{2}\int_S B(\nabla^2 z)^2 ds \ . \qquad (2)$$

Here, $B = \dfrac{Eh^3}{12(1-v^2)}$ is bending stiffness, $E$ is the Young's modulus, $h$ is the thickness, and $v$ is Poisson's ratio of graphene [24,39]. We define a characteristic area $S = \lambda L$ of a single wrinklon. By replacing $\lambda'$ with $L$ in Eq. (2), we obtain

$$U_B = \frac{Eh^3 A^2}{24(1-v^2)}\left[\left(\frac{2\pi}{\lambda}\right)^4 + \left(\frac{2\pi}{L}\right)^4\right]\int_0^\lambda \sin^2\left(\frac{2\pi}{\lambda}x\right)dx\int_0^L \sin^2\left(\frac{2\pi}{L}y\right)dy$$

$$= \frac{Eh^3 \pi^4 A^2}{6(1-v^2)}\lambda^{-3}L + \frac{Eh^3\pi^4 A^2}{6(1-v^2)}\lambda L^{-3} \ . \qquad (3)$$

Here, the first and second terms are the bending energy in the $x$ and $y$ directions, i.e., $U_{Bx}$, and $U_{By}$, respectively. Obviously, $U_{Bx}/U_{By} = L^4/\lambda^4 \gg 1$. Therefore, the bending energy in the $y$ direction is neglectable in our system. This is in accordance with our experimental result that we only observe apparent periodic wrinkles along the $x$ axis.

Next, we will focus on the stretching energy of the graphene flakes. The relative change of the length of the graphene sheet along the $x$ direction results in the periodic wrinkles and

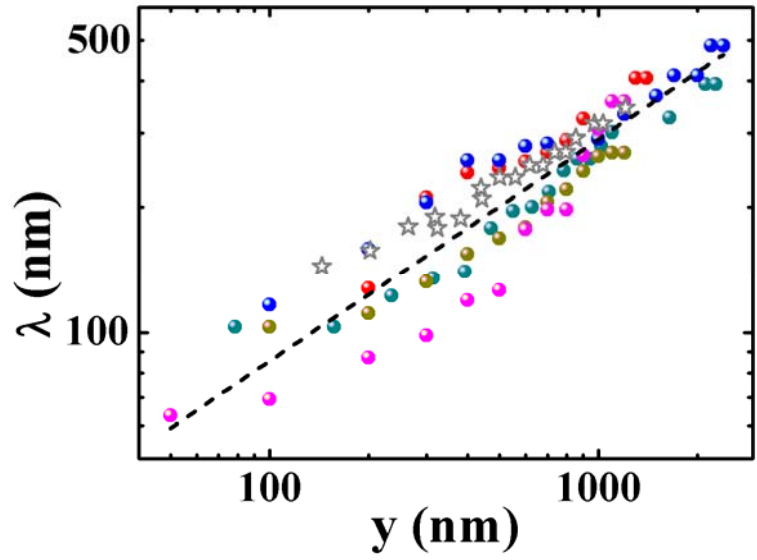

FIG. 5 (color online). Evolution of the average wavelength λ with the distance to the edge y for the hexagonal graphene flakes on Cu surface. The gray stars are the experiment data of strained graphene bilayer reproduced from Ref. [23]. The black dashed curve is for the fitting result by $\lambda \approx (ky)^m$ with $k \approx$ 55 nm and $m \approx 0.5$.

therefore contributes mainly to the bending energy. It is reasonable to neglect the stretching energy in the $x$ direction. The temperature dependent thermal expansion coefficient of graphene generates an effective force $F = Eh\int_{T_0}^{T_1}\alpha_G(T)dT$ per unit length on graphene [40]. Then, the stretching energy of the graphene flake in the $y$ direction can be written as

$$U_S \approx U_{Sy} = \frac{1}{2}\int_S (\partial_y z)^2 dS \int_{T_0}^{T_1} Eh\alpha_G(T)dT$$

$$= \frac{1}{2} Eh\pi^2 A^2 \Delta_y \lambda L^{-1} \quad . \tag{4}$$

Consequently, the total energy of a wrinklon in the graphene flake is given by

$$U_{tot} = \frac{Eh^3\pi^4 A^2}{6(1-v^2)}\lambda^{-3}L + \frac{Eh\pi^2 A^2 \Delta_y}{2}\lambda L^{-1} \quad . \tag{5}$$

Minimizing the total energy $U_{tot}$ with respect to $L$ yields

$$L = \frac{[3\Delta_y(1-v^2)]^{1/2}}{h\pi}\lambda^2 \quad . \tag{6}$$

The obtained relation between the length $L$ and the wavelength $\lambda$ reflects a balance between the bending and stretching energies of graphene flakes. At the boundaries of graphene, the constraint, combined with the tendency to increase the wrinkling wavelength (the tendency to reduce the bending energy), leads to the hierarchical wrinkling pattern. Then, the spatial evolution of the wavelength of the hierarchical patterns in the hexagonal graphene sheets can be obtained by integration of Eq. (1) with $L(\lambda)$, and we obtain:

$$\lambda = \left[ \frac{2\pi h}{\sqrt{3\Delta_y(1-v^2)}} y \right]^{0.5}. \tag{7}$$

In the calculation, we used a boundary condition: $\lambda = \lambda_0 \sim 0$ at $y = 0$. The exponent $m = 0.5$ is in very good agreement with the experimental data. The parameter $k = \frac{2\pi h}{\sqrt{3\Delta_y(1-v^2)}}$ [41] in Eq. (7) reflects the vertical intercept in Fig. 5. While, the relative length change of graphene $\Delta_y = \int_{T_0}^{T_1} \alpha_G(T)dT$ depends on the variations in temperature. In our experiment, the relative length change of graphene $\Delta_y$ is expected to be a constant. For the thermal strain engineering used in Ref. [23], it is expected to have similar value of $\Delta_y$ according to its expression. Therefore, it is reasonable to obtain almost identical evolution of the wavelength $\lambda$ with the distance to the edge as that reported in Ref. [23]. A tiny offset in vertical axis, as shown in Fig. 5, may arise from the different thickness of graphene monolayer and bilayer. Therefore, the Eq. (7) depicts a very robust feature of hierarchical wrinkles in graphene system induced by thermal manipulation. For graphene monolayer, there is the ambiguity of defining the effective thickness for the single layer of C atoms [17,42], and $h = 0.335$ nm (the experimentally measured interlayer spacing in graphite), is widespread used in many theoretical and experimental works. According to the experimental data shown in Figure 5, $\Delta_y$ is estimated as 0.051% with the thickness of graphene $h \approx 0.335$ nm and the Poisson's ratio of graphene $v \approx 0.165$. The obtained small value of $\Delta_y$ is well consistent with the fact that the thermal expansion coefficient of graphene is extremely small and, additionally, the thermal expansion coefficient of graphene changes signs in the studied temperature range of the thermal manipulation [37,38].

In summary, we systematically studied the hierarchy of graphene wrinkles induced by thermal strain engineering and demonstrated that the hierarchical patterns in the graphene flakes can be described quite well by the equation $\lambda \approx (ky)^{0.5}$. Our work demonstrates a universal self-similar hierarchical patterns in this ultimate thin film, and further points out that the classical membrane behavior of graphene persists down to about 100 nm of the wrinkling wavelength.

**Acknowledgements**

We are grateful to National Key Basic Research Program of China (Grant No. 2014CB920903, No. 2013CBA01603, No. 2013CB921701), National Science Foundation (Grant No. 11374035, No. 11004010, No. 10974019, No. 51172029, No. 91121012), and the Fundamental Research Funds for the Central Universities. Lan Meng and Ying Su contributed equally to this paper.

*corresponding author: helin@bnu.edu.cn .